\documentclass[twocolumn,showpacs,superscriptaddress,preprintnumbers,amsmath,amssymb,footinbib]{revtex4}
\newcommand{\lsim}{\mathrel{\rlap{\raisebox{.3ex}{$<$}}
    \raisebox{-.6ex}{$\sim$}}}

\usepackage{url}
\usepackage{graphicx}
\begin{document}

\title{\bf Prospects for measuring coherent neutrino-nucleus elastic scattering at a stopped-pion neutrino source}
\author{Kate Scholberg}\affiliation{Department of Physics, Duke University, Durham, NC 27708 USA}
\date{\today}

\begin{abstract}
Rates of coherent neutrino-nucleus elastic scattering at a
high-intensity 
stopped-pion neutrino source 
in various detector materials (relevant for novel low-threshold
detectors) are calculated.  Sensitivity of a coherent neutrino-nucleus elastic
scattering experiment to new physics is also explored.
\end{abstract}

\maketitle

\section{Introduction}\label{intro}

Coherent elastic neutral current (NC) neutrino-nucleus
scattering~\cite{Freedman:1977xn, Drukier:1983gj} has
never been observed.  In this process, a neutrino of
any flavor scatters off a nucleus at low momentum transfer $Q$ such
that the nucleon wavefunction amplitudes are in phase and add
coherently.  The cross section for a spin-zero nucleus, neglecting
radiative corrections, is given by~\cite{Horowitz:2003cz},

\begin{equation}
\left(\frac{d\sigma}{dE}\right)_{\nu A} = \frac{G_F^2}{2\pi} \frac{Q_w^2}{4} F^2(2ME) M \left[2 - \frac{M E}{k^2}\right], 
\end{equation}
where $k$ is the incident neutrino energy, $E$ is the nuclear recoil
energy, $M$ is the nuclear mass, $F$ is the ground state elastic form
factor, $Q_w$ is the weak nuclear charge, and $G_F$ is the Fermi
constant. The condition for coherence requires that $Q\lsim \frac{1}{R}$,
where $R$ is the nuclear radius. 
This condition is largely satisfied for neutrino energies up to $\sim$50~MeV
for medium $A$ nuclei~\cite{Drukier:1983gj, BoehmandVogel}.

For neutrino energies up to $\sim$50~MeV, typical values of the total
coherent elastic cross section are in the range $\sim
10^{-39}$~cm$^2$, which is relatively high compared to other neutrino
interactions in that energy range (e.g. charged current (CC)
inverse beta decay on protons has
a cross section $\sigma_{\bar{\nu}_e p}\sim 10^{-40}$~cm$^2$, and elastic
neutrino-electron scattering has a cross section\
$\sigma_{\nu_e e}\sim 10^{-43}$~cm$^2$).

In spite of its large cross section, coherent elastic neutrino-nucleus
scattering has been difficult to observe due to the very small
resulting nuclear recoil energies: the maximum recoil energy is 
$\sim 2 k^2/M$, which is in the sub-MeV range for $k~\sim~50$~MeV for
typical detector materials (carbon, oxygen).  Such energies are below
the detection thresholds of most conventional high-mass neutrino
detectors. Although there have been suggestions to look for coherent elastic
$\nu A$ scattering of reactor, spallation source, solar, supernova, or
geophysical
neutrinos~\cite{Drukier:1983gj,Barbeau:2002fg,Hagmann:2004pb,Wong:2005vg},
as yet there has been no successful detection.

However, in recent years there has been a surge of progress in
development of novel ultra low threshold detectors, many with the aim
of weakly interacting massive particle
recoil detection or very low energy solar neutrino detection.  Thresholds
of 10~keV or even lower for detection of nuclear recoils may be
possible.  Such detectors include (but are not limited to): noble element
(neon, argon, xenon) scintillation, ionization or tracking detectors,
solid state detectors (germanium, silicon), bubble and superheated droplet
detectors~\cite{McKinsey:2004rk,
Boulay:2004nx,Boulay:2004dk,Aprile:2005mt,Takeuchi:2004df,Alner:2005pa,Snowden-Ifft:2004sz,Barbeau:2002fg,Hagmann:2004pb,Galeatalk,
Akerib:2004iz, Akerib:2004in,Bolte:2005fm,Barnabe-Heider:2005ri}
. Some of these new technologies (for instance noble liquid detectors
such as CLEAN~\cite{McKinsey:2004rk}) may plausibly attain ton-scale
masses in the relatively near future.

\begin{figure}[!ht]
\begin{centering}
\includegraphics[height=2in]{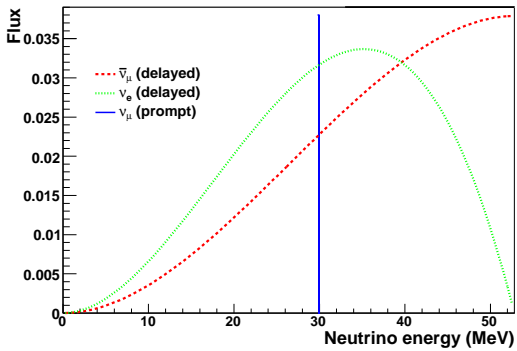}
\caption{Shape of neutrino spectra from a stopped-pion source, for
the different produced flavors.}
\label{fig:sns_spec}
\end{centering}
\end{figure}

A promising source of neutrinos for measurement of coherent elastic
cross sections is that arising from a stopped-pion source.
Monoenergetic 29.9~MeV $\nu_{\mu}$'s are produced from pion decay at
rest, $\pi^{+}~\rightarrow~ \mu^{+}~\nu_\mu$, and
$\bar{\nu}_\mu$ and $\nu_e$ from
$\mu^{+}~\rightarrow~\nu_e~e^{+}~\bar{\nu}_{\mu}$ follow on a
muon-decay timescale ($\tau=2.2~\mu$s).  The neutrino spectral shape
is shown in Fig.~\ref{fig:sns_spec}.  Neutrinos in this energy range
will produce nuclear recoils from coherent scattering with tens of
keV.  If the beam is pulsed in a short ($< \mu$s) time window, the
pion decay $\nu_{\mu}$ will be prompt with the beam, and the muon-decay 
products will be delayed.

Stopped-pion sources of neutrinos have been employed for neutrino
experiments in
the past~\cite{Athanassopoulos:1996ds, Burman:1996gt}, and future high-flux facilities are 
planned~\cite{Gardner:1998bu}. The Spallation Neutron Source (SNS) at
Oak Ridge National Laboratory should turn on in 2006; J-PARC is
another near-term possibility~\cite{jparc}.  An experiment to measure
neutrino-nucleus CC and NC cross sections 
(relevant for supernova physics and
detection) in the tens of MeV
range~\cite{Avignone:2003ep,nusns} with conventional detectors at the SNS has already been proposed.  A
shielded concrete bunker is envisioned at a location 20~m from the
source; this could potentially accommodate a low-threshold detector
as well as the currently planned detectors.
The SNS beam is pulsed, with less than microsecond width and 
a 60~Hz frequency.

Here prospects for measuring coherent elastic neutrino-nucleus
scattering will be evaluated
using parameters relevant for the SNS;
however the results should be generally applicable to experiments at
any high-intensity stopped-pion $\nu$ source.

\section{Expected Event Rates}

\begin{figure}[!htbp]
\begin{centering}
\includegraphics[height=2.8in]{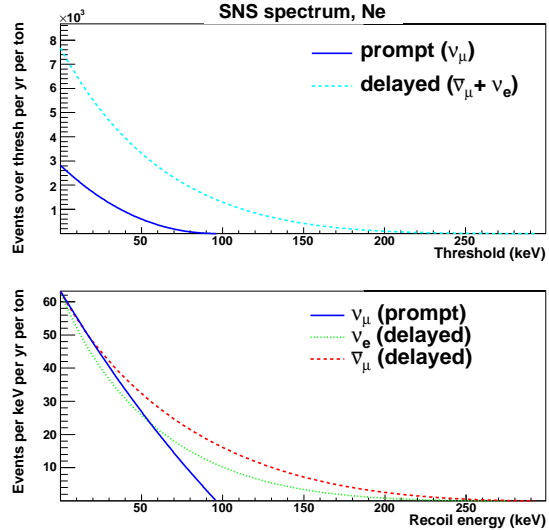}
\caption{Bottom panel: Differential yield at the SNS in 1 ton of $^{20}$Ne
(solid: $\nu_\mu$, dotted: $\nu_e$, dashed: $\bar{\nu}_\mu$) per year per keV,
as a function of recoil energy.
Top panel: Number of interactions over recoil energy threshold
in 1 ton of $^{20}$Ne for 1 yr of running at the SNS
(solid: $\nu_\mu$, dashed: sum of $\nu_e$ and $\bar{\nu}_\mu$),
as a function of recoil energy threshold.}
\label{fig:snsyield_ne}
\end{centering}
\end{figure}

\begin{figure}[!htbp]
\begin{centering}
\includegraphics[height=2.8in]{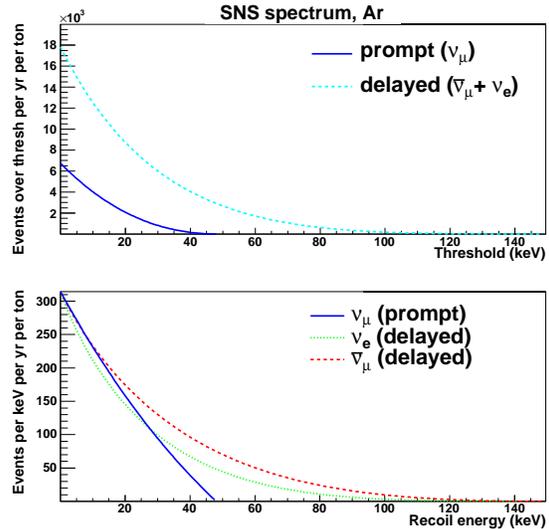}
\caption{As in Fig.~\ref{fig:snsyield_ne} for $^{40}$Ar.}
\label{fig:snsyield_ar}
\end{centering}
\end{figure}

\begin{figure}[!htbp]
\begin{centering}
\includegraphics[height=2.8in]{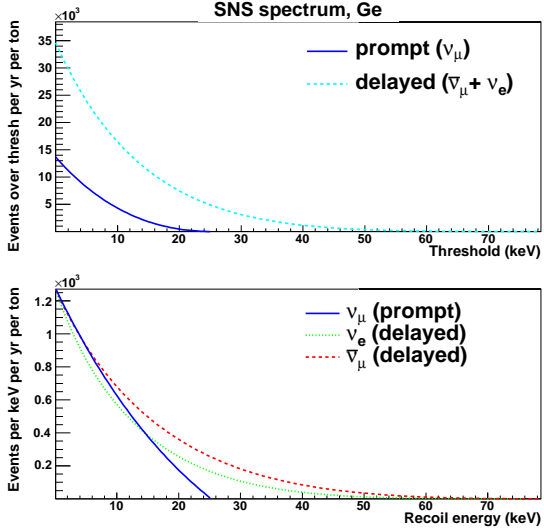}
\caption{As in Fig.~\ref{fig:snsyield_ne} for $^{76}$Ge.}
\label{fig:snsyield_ge}
\end{centering}
\end{figure}

\begin{figure}[!htbp]
\begin{centering}
\includegraphics[height=2.8in]{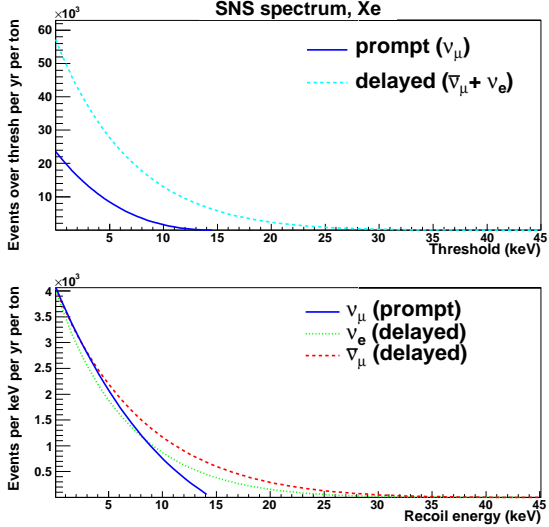}
\caption{As in Fig.~\ref{fig:snsyield_ne} for $^{132}$Xe.}
\label{fig:snsyield_xe}
\end{centering}
\end{figure}

The expected rate of interactions differential in recoil energy is given by
\begin{equation}
\frac{dN}{dE} = N_t \int dk \phi(k) \frac{d\sigma}{dE}(k),
\end{equation}
where $N_t$ is the number of targets and $\phi(k)$ is the incident neutrino
flux.
Spectra for $\nu_\mu$, $\bar{\nu}_\mu$ and
$\nu_e$ for a stopped $\pi^+/\mu^+$ source, 
assuming $\sim 10^7$ $\nu$ s$^{-1}$ cm$^{-2}$ per flavor at
20~m from the source are assumed.  Cross sections and form factors
from~\cite{Horowitz:2003cz, horowitz2} for $^{20}$Ne, $^{40}$Ar,
$^{76}$Ge, and $^{132}$Xe are used.
Figs.~\ref{fig:snsyield_ne} through~\ref{fig:snsyield_xe} show the results. The rates are quite
promising: for a ton-scale detector with a few to 10~keV threshold,
$10^{4}-10^{5}$ signal events per year are expected.  Even for
kilogram-scale detectors, event rates may be in the tens per year.

\begin{figure}[!htbp]
\begin{centering}
\includegraphics[width=3.2in]{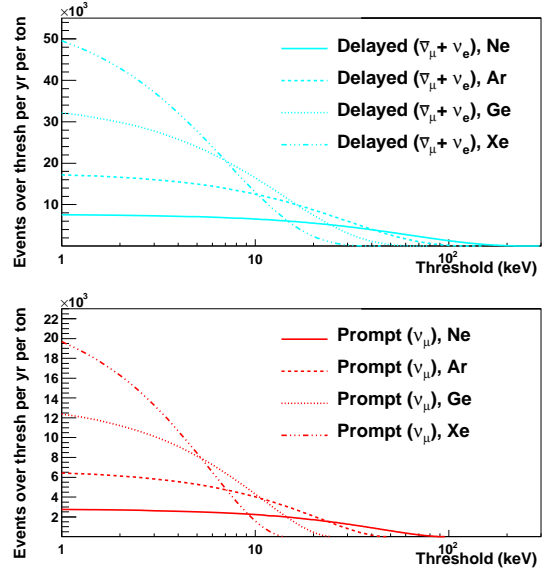}
\caption{The number of interactions over the recoil energy threshold
for various detector materials
(bottom panel: prompt $\nu_\mu$, top panel: sum of delayed $\nu_e$ and $\bar{\nu}_\mu$).}
\label{fig:compare_yield}
\end{centering}
\end{figure}

Fig.~\ref{fig:compare_yield} plots integrated yield over threshold
for several elements for comparison.  One can see that the higher the
nuclear mass, the higher the overall event rate at low threshold
(scaling approximately as the square of the number of neutrons), but
the smaller the typical recoil energies ($E_{{\rm max}}= 2 k^2/M$).

In the absence of a specific detector model,
perfect detection efficiency and zero background are assumed,
which is not realistic.
Detection
efficiencies for many
low-threshold detector types (see Sec.~\ref{intro}) can
be reasonably high, but can depend on background levels.
Backgrounds will include
beam-related neutrons, cosmogenics, radioactivity and
instrumental background, as well as other CC and NC neutrino
reactions; these will need to be evaluated for a specific
detector's rejection capabilities and location.  
Backgrounds are not obviously overwhelming, especially
given that the pulsed structure of the beam such
as that at the SNS leads to a powerful
rejection factor ($\sim 4 \times 10^{-4}$) against steady-state
backgrounds.
It is not really clear at this point
whether beam-related backgrounds will be worse for prompt or delayed
neutrinos; it will depend on shielding and detector location. Therefore
the contributions from prompt and delayed fluxes are given separately.

\section{Physics Potential}

The neutrino-nucleus coherent elastic scattering cross section is
cleanly predicted by the Standard Model (SM); form factors can be
known to better than 5\%, and radiative corrections are known at the percent
level~\cite{horowitz3}.  Any measured deviations from prediction would
be interesting~\cite{Krauss:1991ba}.  In the context of the SM, the
weak mixing angle is related to the absolute scattering rate.  One can
also constrain non-standard interactions (NSI) of neutrinos and
nucleons. Also, non-zero neutrino magnetic moment will modify the
cross section at low energies.  There are further reasons to measure
coherent neutrino-nucleus scattering: neutrino-nucleus scattering
processes are important in supernova physics~\cite{Freedman:1977xn},
as well as being useful themselves for supernova neutrino
detection~\cite{Horowitz:2003cz}.  Because they are flavor blind, NC
processes allow measurement of total neutrino flux, which can be
compared to independently measured CC interactions.  Therefore one can
obtain limits on neutrino oscillation, and in particular on
oscillations to sterile neutrinos~\cite{note1}.
Finally, it has even been proposed to
exploit the large cross sections of neutrino-nucleus scattering for
practical neutrino detectors, e.g. reactor
monitoring~\cite{Barbeau:2002fg,Hagmann:2004pb}.

This section will discuss the various ways of probing new physics with
a coherent elastic scattering experiment.  At this stage, the
experimental systematic uncertainty on the absolute rate is not
known. It will depend on the specific detector type and configuration,
backgrounds, and source uncertainties. However a total systematic
uncertainty of $\sim$10\% (including nuclear, beam and
detector-related uncertainties), while perhaps optimistic, may well be
achievable.  Systematic uncertainties will likely dominate at the few
tens of a kilogram scale or greater.

\subsection{Weak Mixing Angle}

The SM predicts a coherent elastic scattering rate proportional
to $Q_w^2$, the weak charge given by $Q_w = N-(1-4\sin^2 \theta_W)Z$,
where $Z$ is the number of protons, $N$ is the number of neutrons
and $\theta_W$ is the weak mixing angle.
Therefore the weak mixing angle can be extracted from the
measured absolute cross section, at a typical $Q$ value of 0.04~GeV/$c^2$.
A deviation from the SM prediction could indicate new physics.

If the absolute cross section can be measured to 10\%, there
will be an uncertainty on $\sin^2 \theta_W$ of $\sim 5 \%$.  This
is not competitive with the current best measurements from
atomic parity violation~\cite{Bennett:1999pd,Eidelman:2004wy}, 
SLAC E158~\cite{Anthony:2005pm} and NuTeV~\cite{Zeller:2001hh},
which have better than percent-level uncertainties.  One would
need to significantly improve
the systematic uncertainty on the absolute rate
(perhaps by normalizing with 
a well-known rate) for coherent elastic $\nu$A scattering 
in order to make a useful measurement of the weak mixing angle.  More promising
would be a search for non-standard interactions of neutrinos
with nuclei, as described in the following subsection.

\subsection{Non-Standard Interactions of Neutrinos}

Existing precision measurements of the weak mixing angle at low $Q$ do not
constrain new physics which is specific to neutrino-nucleon interactions.

Here a model-independent parameterization of
non-standard contributions to the cross section is used, following
Refs.~\cite{Barranco:2005yy, Davidson:2003ha}.  
In this description, one assumes an effective Lagrangian for interaction
of a neutrino with a hadron:

\begin{eqnarray}
\lefteqn{ \mathcal{L}^{NSI}_{\nu H}} & =  - \frac{G_F}{\sqrt{2}}{\displaystyle \sum_{{q=u,d}\atop{\alpha, \beta = e, \mu, \tau}}} [ \bar{\nu}_\alpha \gamma^\mu (1-\gamma^5) \nu_\beta]\times & \\ \nonumber
& (\varepsilon_{\alpha \beta}^{qL}[ \bar{q} \gamma_\mu (1-\gamma^5) q]+ \varepsilon_{\alpha \beta}^{qR} [ \bar{q} \gamma_\mu (1+\gamma^5) q]).  &
 \end{eqnarray}
The $\varepsilon$ parameters describe either ``non-universal'' ($\alpha = \beta$)
or flavor-changing ($\alpha \ne \beta$) interactions.  

As in Ref.~\cite{Barranco:2005yy}, nuclei with total spin zero,
and for which sum of proton spins and sum of neutron spins is also
zero, are considered; in this case we have sensitivity to vector
couplings, $\varepsilon_{\alpha \beta}^{qV}= \varepsilon_{\alpha
\beta}^{qL}+\varepsilon_{\alpha \beta}^{qR}$.  The cross section for
coherent NC elastic scattering of neutrinos of flavor $\alpha$ off
such a spin-zero nucleus is given by

\begin{eqnarray}\label{eq:xscn}
\lefteqn{\left(\frac{d\sigma}{dE}\right)_{\nu_\alpha A} = 
  \frac{G_F^2 M}{\pi} F^2(2ME)\left[1 - \frac{M E}{2k^2}\right] \times} \\
&     \{[Z(g_V^p + 2\varepsilon_{\alpha \alpha}^{uV}+ \varepsilon_{\alpha \alpha}^{dV})+ N(g_V^{n} + \varepsilon_{\alpha \alpha}^{uV}+ 2\varepsilon_{\alpha \alpha}^{dV})]^2 & \nonumber \\ 
 &            + \displaystyle \sum_{\alpha \ne \beta }{[Z(2\varepsilon_{\alpha \beta}^{uV}+ \varepsilon_{\alpha \beta}^{dV})+ N(\varepsilon_{\alpha \beta}^{uV} + 2 \varepsilon_{\alpha \beta}^{dV})]^2 \}}, & \nonumber 
 \end{eqnarray}
where $Z$ is the number of protons in the nucleus, $N$ is the number
of neutrons, and $g_V^p = (\frac{1}{2}-2\sin^2 \theta_W)$, $g_V^n=-\frac{1}{2}$ are the SM weak constants.

A stopped-pion neutrino
source such as that at the SNS contains $\nu_\mu$, $\bar{\nu}_\mu$, and $\nu_e$.
A coherent elastic $\nu$A scattering experiment employing
such a source would therefore have sensitivity to all but
$\varepsilon_{\tau \tau}$ couplings.

Existing constraints on the values of $\varepsilon_{\alpha \beta}^{P}$
($P=L,R$) are summarized in Ref.~\cite{Davidson:2003ha}. 
Table~\ref{tab:constraints} selects those relevant for interactions
of electron and muon flavor neutrinos with quarks.
New constraints from existing
and future atmospheric, beam and solar neutrino experiments are explored
in Refs.~\cite{Friedland:2005vy,Friedland:2004pp}.

\begin{table}[h]
\begin{centering}
\caption{Constraints on NSI parameters, from Ref.~\cite{Davidson:2003ha}.}
\label{tab:constraints}
\begin{tabular}{cc}\hline\hline
NSI Parameter Limit & Source \\ \hline
$-1 < \varepsilon^{uL}_{ee} < 0.3$  &  CHARM $\nu_eN$, $\bar{\nu}_eN$ scattering\\ 
$-0.4<\varepsilon^{uR}_{ee} < 0.7$   & \\ 

$-0.3 < \varepsilon^{dL}_{ee} < 0.3$   &  CHARM $\nu_eN$, $\bar{\nu}_eN$ scattering \\ 
$-0.6<\varepsilon^{dR}_{ee} < 0.5$   &  \\ 

$|\varepsilon^{uL}_{\mu \mu}| < 0.003$   &  NuTeV $\nu N$, $\bar{\nu}N$ scattering\\ 
$-0.008<\varepsilon^{uR}_{\mu \mu} < 0.003$   &\\ 

$|\varepsilon^{dL}_{\mu \mu}| < 0.003$  &  NuTeV $\nu N$, $\bar{\nu} N$ scattering \\ 
$-0.008<\varepsilon^{dR}_{\mu \mu} < 0.015$   & \\ 

$|\varepsilon^{uP}_{e\mu}| < 7.7\times 10^{-4}$   &  $\mu\rightarrow e$ conversion on nuclei \\ 

$|\varepsilon^{dP}_{e\mu}| < 7.7\times 10^{-4}$   &  $\mu\rightarrow e$ conversion on nuclei \\ 

$|\varepsilon^{uP}_{e \tau}| < 0.5$   &  CHARM $\nu_eN$, $\bar{\nu}_eN$ scattering\\ 

$|\varepsilon^{dP}_{e \tau}| < 0.5$   &  CHARM $\nu_eN$, $\bar{\nu}_eN$ scattering\\ 

$|\varepsilon^{uP}_{\mu \tau}| < 0.05$   &  NuTeV $\nu N$, $\bar{\nu}N$ scattering\\ 

$|\varepsilon^{dP}_{\mu \tau}| < 0.05$   &  NuTeV $\nu N$, $\bar{\nu}N$ scattering\\ 

\hline \hline

\end{tabular}
\end{centering}
\end{table}

From this table, one can see that of these parameters,
$\varepsilon_{ee}$ and $\varepsilon_{e\tau}$ are quite poorly
constrained: values of order unity are allowed.  $|\varepsilon_{\mu
\beta}|$ couplings are, however, constrained to better than 0.05.
Given this situation, the focus here is on $\varepsilon_{ee}$ and
$\varepsilon_{e \tau}$ 
couplings~\cite{note2}.
These would be accessible using the electron flavor
component of the source.
That no oscillations take place (i.e. that the standard
three-flavor model of neutrino mixing holds, and that the
baseline is too short for significant flavor transition) is also assumed. 

The signature of NSI is a deviation from the expected cross section.
The following show a few examples of two-dimensional slices of regions
in $\varepsilon_{\alpha \beta}$ parameter space that would be allowed if
one measured exactly the SM expectation.

\begin{figure}[!ht]
\begin{centering}
\includegraphics[height=2.8in]{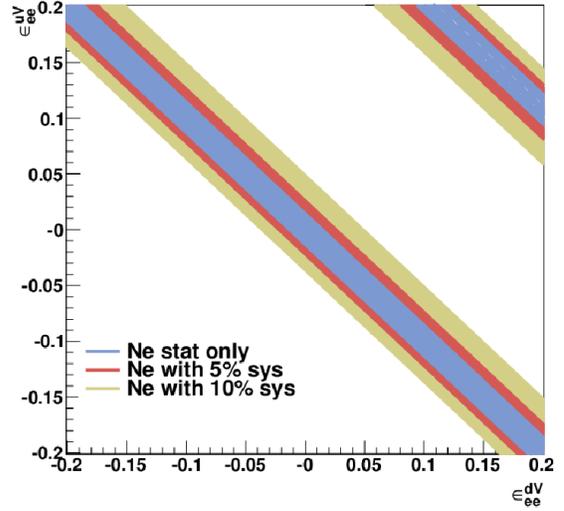}
\caption{Allowed region at 90\% C.L. for $\varepsilon_{ee}^{uV}$ and
$\varepsilon_{ee}^{dV}$, for 100 kg-yr of $^{20}$Ne at the SNS.  The outer
region corresponds to an assumed systematic uncertainty 
of 10\% in addition to statistical uncertainty; 
the middle region corresponds to an
assumed systematic uncertainty of 5\%, and the inner region corresponds
to statistical uncertainty only.}
\label{fig:Ne_ee}
\end{centering}
\end{figure}

Fig.~\ref{fig:Ne_ee} shows 90\% C. L. allowed regions one would draw
for $\varepsilon_{ee}^{uV}$, $\varepsilon_{ee}^{dV}$, if the rate
predicted by the SM were measured for the delayed flux (which contains
$\nu_e$), assuming that the $\varepsilon_{\mu \beta}$ parameters are
negligible, and for $\varepsilon_{e\tau}^{qV} = 0$, for 100 kg-yr of
running of a neon detector at 20~m from the source.  A 10~keV
threshold is assumed.  This calculation considers only the total
delayed ($\nu_e+\bar{\nu}_\mu$) flux rate~\cite{note3}.
The regions corresponding to
assumptions of 5\% and 10\% systematic error in addition to
statistical error, and for statistical error alone are
shown~\cite{note4}.
As before, a perfectly efficient,
background-free detector is assumed.

Note that in Eq.~\ref{eq:xscn}, even in the presence of non-universal
NSI, one can obtain
rates identical to the SM prediction in the case
that 
\begin{eqnarray}
\lefteqn{Z(g_V^p + 2 \varepsilon_{ee}^{uV} + \varepsilon_{ee}^{dV})
   + N(g_V^n + \varepsilon_{ee}^{uV} +  2 \varepsilon_{ee}^{dV})} \hspace{1.5in}  \\
   &  & = \pm (Z g_V^p + N g_V^n), \nonumber
\end{eqnarray}
so for 
\begin{equation*}
\varepsilon_{ee}^{uV} = -\frac{(A+N)}{(A+Z)} \varepsilon_{ee}^{dV},
\end{equation*}
and
\begin{equation*}
\varepsilon_{ee}^{uV} = -\frac{(A+N)}{(A+Z)} \varepsilon_{ee}^{dV}-\frac{2(Z g_V^p + N g_V^n)}{A+Z}. 
\end{equation*}
For this reason, allowed regions of Fig.~\ref{fig:Ne_ee}
appear as linear bands in $\varepsilon_{ee}^{uV}$,
$\varepsilon_{ee}^{dV}$ parameter space.
A measurement employing more than one
element can then place more stringent constraints on the couplings; 
the more 
the $(A+N)/(A+Z)$ ratio differs between the two targets, the better. 

\begin{figure}[!ht]
\begin{centering}
\includegraphics[height=2.8in]{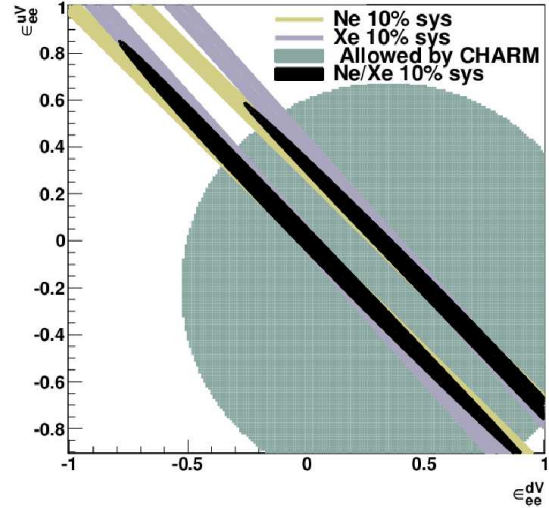}
\caption{Allowed regions at 90\% C.L. for $\varepsilon_{ee}^{uV}$ and
$\varepsilon_{ee}^{dV}$, for 100 kg-yr each of $^{20}$Ne
and $^{132}$Xe (steeper slope band) at the SNS, assuming 10\% systematic uncertainty, plus
statistical uncertainty.
The thin black ellipses correspond to
combined Ne/Xe measurement.  The shaded 
elliptical region corresponds to a slice of the
CHARM experiment's allowed NSI parameter space, for
$\varepsilon_{ee}^{qA}=0$.}
\label{fig:NeXe_ee}
\end{centering}
\end{figure}

Fig.~\ref{fig:NeXe_ee} shows the same regions for 100 kg-yr each
of $^{132}$Xe and $^{20}$Ne, where the black ellipses represent the 90\% allowed
region from the combination of the measurements.  These regions are
superposed on the allowed region from high-energy $\nu_e$ scattering
on nucleons derived from CHARM experiment results~\cite{Dorenbosch:1986tb} in
Ref.~\cite{Davidson:2003ha}, for the case that the axial
parameters
$\varepsilon_{ee}^{qA}=\varepsilon_{ee}^{qL}-\varepsilon_{ee}^{qR}$
are zero.

\begin{figure}[!ht]
\begin{centering}
\includegraphics[height=2.8in]{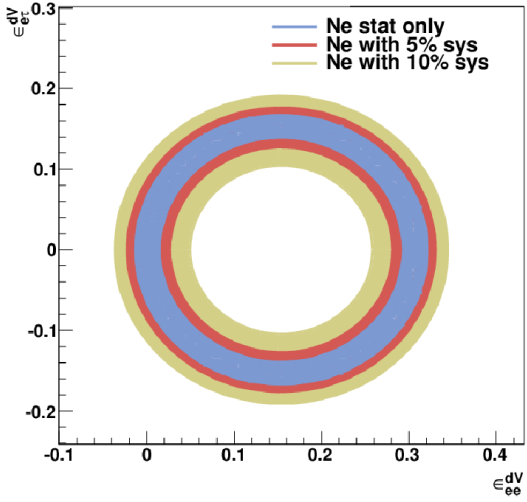}
\caption{Allowed region at 90\% C.L. for $\varepsilon_{ee}^{dV}$ and
$\varepsilon_{e\tau}^{dV}$, for 100 kg-yr of $^{20}$Ne at the SNS. The
shaded region between the outer and inner ellipses
corresponds to an assumed systematic uncertainty 
of 10\% in addition to statistical uncertainty; 
the next largest region corresponds to an
assumed systematic uncertainty of 5\%, and the inner region corresponds
to statistical uncertainty only.}
\label{fig:Ne_etau}
\end{centering}
\end{figure}

\begin{figure}[!ht]
\begin{centering}
\includegraphics[height=2.8in]{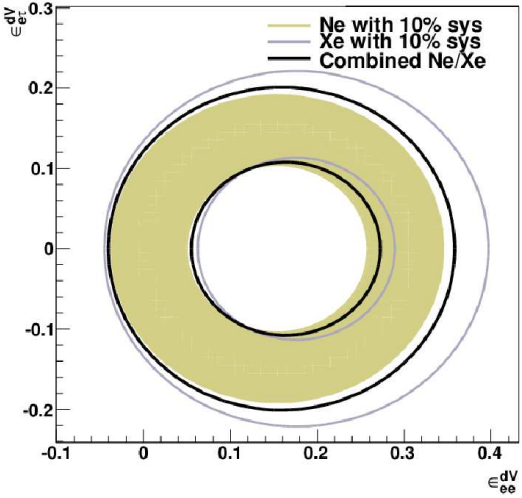}
\caption{
Allowed regions with same assumptions
as Fig.~\ref{fig:Ne_etau}, 10\% systematic uncertainty,
for $^{20}$Ne (shaded region), $^{132}$Xe (region between pale lines)
and both combined (region between black lines).}
\label{fig:Ne_etau2}
\end{centering}
\end{figure}

Fig.~\ref{fig:Ne_etau} shows similar 90\% allowed regions for a
slice of $\varepsilon_{ee}^{dV}$, $\varepsilon_{e\tau}^{dV}$ parameter
space, for $\varepsilon_{ee}^{uV}=\varepsilon_{e\tau}^{uV}=0$ (note
that $d$-quark NSI may be especially interesting; see
e.g.~\cite{Amanik:2004vm}). In this case the allowed parameters
correspond to regions between two ellipses.
Fig.~\ref{fig:Ne_etau} shows regions for a $^{20}$Ne detector (with same
assumptions as above).  Fig.~\ref{fig:Ne_etau2} shows the result for
$^{132}$Xe as well, and the black ellipses contain the region allowed by the
combined measurements (for this case, only a small improvement is
afforded by measurements with multiple targets).

\begin{figure}[!ht]
\begin{centering}
\includegraphics[height=2.8in]{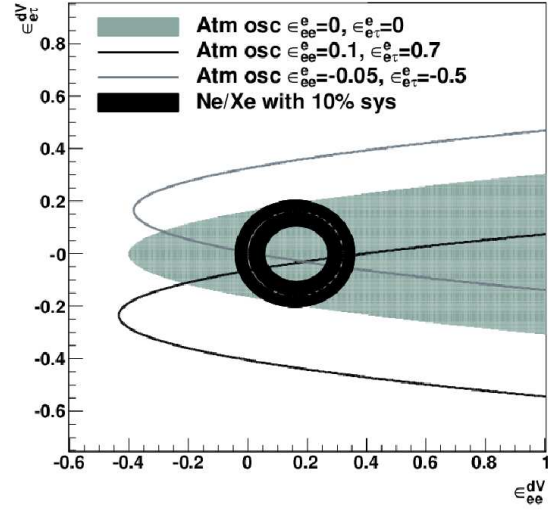}
\caption{Allowed region at 90\% C.L. for $\varepsilon_{ee}^{dV}$ and
$\varepsilon_{e\tau}^{dV}$, $\varepsilon_{ee}^{uV}=\varepsilon_{e \tau}^{uV}=0$,
for 100 kg-yr each of $^{20}$Ne and $^{132}$Xe at the SNS
is shown in black.  The shaded
region corresponds to a slice of allowed NSI parameters 
from Ref.~\cite{Friedland:2005vy} for $\varepsilon_{\tau \tau}=0$,
and $\varepsilon_{ee}^{eV}=\varepsilon_{e \tau}^{eV}=0$; the parabolic
regions inside the dark and light lines correspond to slices of
allowed parameter space for some specific values of 
$\varepsilon_{ee}^{eV}$ and $\varepsilon_{e \tau}^{eV}$.}
\label{fig:NeXe_etau_compare}
\end{centering}
\end{figure}

Fig.~\ref{fig:NeXe_etau_compare} compares neutrino-nucleus scattering 
sensitivity to allowed NSI parameters derived based on lack of
distortion of oscillation parameters for beam and atmospheric
neutrinos~\cite{Friedland:2005vy}.

\begin{figure}[!ht]
\begin{centering}
\includegraphics[height=2.8in]{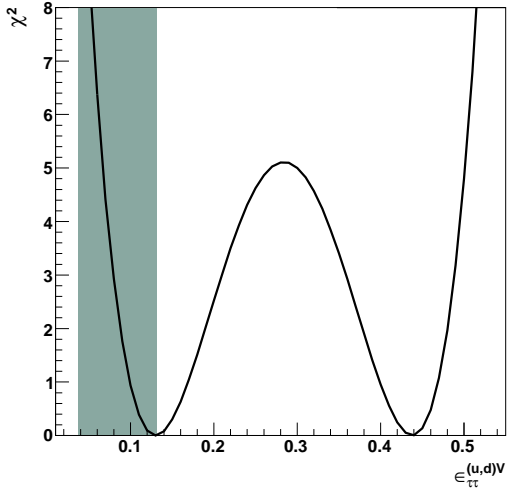}
\caption{$\chi^2$ as a function of $\varepsilon_{\tau \tau}^{(u,d)V}$
for 100 kg-yrs of $^{20}$Ne, assuming NSI parameters 
$\varepsilon_{11}^{(u,d)V}=-0.065$ and $\varepsilon_{12}^{(u,d)V}=-0.15$.
The shaded region represents allowed $\varepsilon_{\tau \tau}$ parameters
from Ref.~\cite{Friedland:2005vy}, from beam and atmospheric neutrino
oscillation constraints.
}
\label{fig:Ne_solar}
\end{centering}
\end{figure}

It is worth noting that a coherent neutrino-nucleus elastic scattering experiment will
provide significant constraints on still-allowed NSI parameters that modify
solar neutrino survival probabilities~\cite{Friedland:2004pp}.  As a
case in point, consider specific NSI parameters that yield the
``LMA-0'' solution of Ref.~\cite{Friedland:2004pp}:
$\varepsilon_{11}^{uV} = \varepsilon_{11}^{dV} = -0.065$, and
$\varepsilon_{12}^{uV} = \varepsilon_{12}^{dV} = -0.15$, where
$\varepsilon_{11} = \varepsilon_{ee} - \varepsilon_{\tau \tau}
\sin^2\theta_{23}$, and $\varepsilon_{12} = -2 \varepsilon_{e\tau}
\sin\theta_{23}$, and $\theta_{23}$ is the atmospheric mixing angle,
known to be $\sim \pi/4$.  Following the approach in this reference, 
$\varepsilon_{\alpha \beta}^{uV}=\varepsilon_{\alpha\beta}^{dV}$ is
assumed.    Fixing
$\varepsilon_{11}$ and $\varepsilon_{12}$ defined in this way will
yield different neutrino-nucleus scattering constraints for different
assumptions of $\varepsilon_{\tau \tau}$.  Fig.~\ref{fig:Ne_solar}
shows the value of a $\chi^2$  defined as
$\chi^2 =  (N_{\rm NSI} - N_{SM})^2/\sigma^2$, as a function of $\varepsilon_{\tau
\tau}^{uV} = \varepsilon_{\tau \tau}^{dV}$;
$N_{\rm NSI}$ is the number of signal events for the given
NSI parameters, and $N_{\rm SM}$ is the SM expectation, for 100 kg-yr of
$^{20}$Ne at 10~keV threshold;
$\sigma^2$ includes both statistical uncertainty
and an assumed 10\% systematic uncertainty.  Superimposed on this plot 
as a shaded region is the restriction on 
$\varepsilon_{\tau\tau}^{(u,d)V}$ (given assumptions above, and
$\varepsilon_{\alpha \beta}^{eV}=0$)
from beam and atmospheric neutrino
oscillations,
$\left|1+\varepsilon_{ee} + \varepsilon_{\tau \tau} - \sqrt{(1+\varepsilon_{ee} - \varepsilon_{\tau \tau})^2 + 4 | \varepsilon_{e\tau}|^2}\right| < 0.4$,
taken from Ref.~\cite{Friedland:2005vy}.

From this plot one can see that 
a coherent elastic neutrino-nucleus scattering experiment has sensitivity to
the set of NSI parameters 
for which the LMA-0 region is derived. If the rate predicted by the SM
were to be observed, these LMA-0 parameters could be
ruled out for $\varepsilon_{\tau \tau} = 0$, and would remain
viable only for a restricted range of $\varepsilon_{\tau \tau}^{(u,d)V}$.

The conclusion of these NSI studies is that
a coherent elastic neutrino-nucleus scattering experiment at a stopped-pion
source would have significant sensitivity to currently-allowed
NSI $\varepsilon_{ee}^{qV}$
and $\varepsilon_{e\tau}^{qV}$ parameters.

\subsection{Neutrino Magnetic Moment}

The SM predicts a neutrino magnetic moment of $\mu_\nu \le
10^{-19} \mu_B (m_{\nu}/1 {\rm eV})$, in units of Bohr
magnetons.  This is very small, but extensions of the SM
commonly predict larger ones.  The most stringent limits are
astrophysical: for instance, based on lack of observed energy
loss from electromagnetic couplings in red giant evolution
one can set a limit $\mu_\nu \le 10^{-12}\mu_B$~\cite{Raffelt:1999gv}.
The best direct experimental limits
result from lack of distortion of neutrino-electron elastic scattering at
low energy, and are in the range of $\mu_\nu(\nu_e) \le 1-2 \times 10^{-10}\mu_
B$~\cite{Liu:2004ny,Daraktchieva:2005kn,Wong:2004sp}.
For muon neutrino scattering, the best limit is less stringent:
$\mu_\nu(\nu_\mu) \le 6.8 \times 10^{-10}\mu_B$~\cite{Auerbach:2001wg}.

A signature of non-zero neutrino magnetic moment can be observed via distortion
of the recoil spectrum of coherently scattered nuclei.
The magnetic scattering cross section is given in
Ref.~\cite{Vogel:1989iv} for a spin-zero nucleus:

\begin{equation}
\left(\frac{d\sigma}{dE}\right)_m = \frac{\pi \alpha^2 \mu_\nu^2 Z^2}{m_e^2}\left(\frac{1-E/k}{E}+ \frac{E}{4k^2} \right).
\end{equation}

\begin{figure}[ht]
\begin{centering}
\includegraphics[width=3.0in]{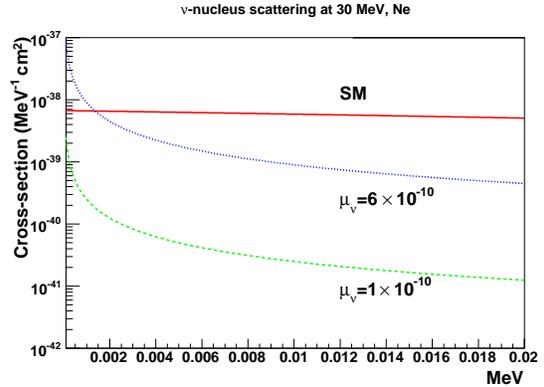}
\caption{Solid line: SM coherent neutrino-nucleus differential
cross section, as a function of nuclear recoil energy $E$, for
neutrino energy $k=30$~MeV and for a $^{20}$Ne target.  Dashed line:
differential cross section for neutrino-nucleus scattering due to a neutrino magnetic
moment of $\mu_\nu = 10^{-10} \mu_B$.  Dotted line: 
differential cross section for neutrino-nucleus scattering due to a neutrino magnetic
moment of $\mu_\nu = 6 \times 10^{-10} \mu_B$.
}
\label{fig:magmom_cohnc}
\end{centering}
\end{figure}

Fig.~\ref{fig:magmom_cohnc} shows the differential cross sections
calculated for $^{20}$Ne, for 30~MeV neutrino energy,
as a function of nuclear recoil energy.  The magnetic
scattering cross section is calculated for neutrino magnetic moment
just below the current best experimental limits ($10^{-10} \mu_B$ for $\nu_e$
and $6\times 10^{-10} \mu_B$ for $\nu_\mu$).

\begin{figure}[ht]
\begin{centering}
\includegraphics[width=3.0in]{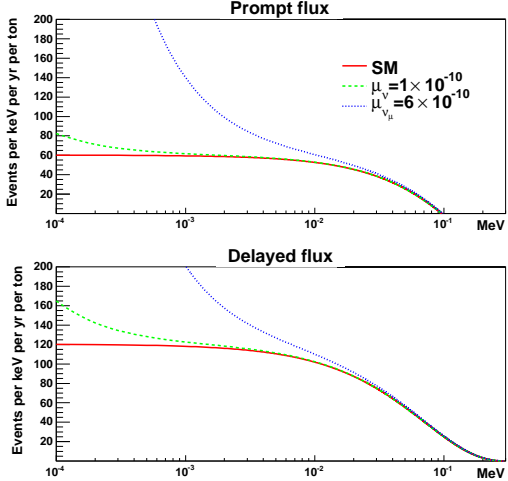}
\caption{Differential yield at the SNS in neon as a function of
nuclear recoil energy. The top plot is for the prompt flux ($\nu_\mu$ only)
and the bottom plot is for the delayed flux (sum of $\nu_e$ and $\bar{\nu}_\mu$).
Solid lines: SM expectation.  Dashed
lines: yield including magnetic moment contribution for $\mu_\nu =
10^{-10} \mu_B$ for both $\nu_e$ and $\bar{\nu}_\mu$. Dotted lines:
yield including magnetic moment contribution for $\mu_\nu =
10^{-10} \mu_B$ for $\nu_e$ and $\mu_\nu =
6 \times 10^{-10} \mu_B$ for $\bar{\nu}_\mu$, $\nu_\mu$.}
\label{fig:magmom_cohnc_yield}
\end{centering}
\end{figure}

Fig.~\ref{fig:magmom_cohnc_yield} shows the yield in events per keV of
recoil energy, per ton per year in a neon detector at $20~$m from
the SNS target, with and without neutrino magnetic
moment contribution, for prompt and delayed fluxes.
The dashed
line assumes $\nu_\mu = 10^{-10} \mu_B$ for both $\nu_e$ and $\bar{\nu}_\mu$.
The dotted line assumes $\nu_\mu = 10^{-10} \mu_B$ for $\nu_e$ and 
$\nu_\mu = 6 \times 10^{-10} \mu_B$ for $\bar{\nu}_\mu$.

The difference in coherent neutrino-nucleus scattering yield due to presence of
a neutrino magnetic moment near the current $\mu_\nu$ limit for $\nu_e$  is very small, except for recoil energies below a few keV.  This signal is therefore
likely out of reach for a CLEAN-type experiment at the SNS.  However,
for $\mu_\nu$ near the current limit for $\nu_\mu$, there might 
be a measurable signal for a 10~keV threshold, and it is conceivable
that one could improve the limit with a high-statistics measurement.
Nuclei with spin, although not considered here, have additional
$\mu_\nu$-dependent terms in their coherent neutrino-nucleus scattering 
cross sections~\cite{Vogel:1989iv} and may be potential targets
for a neutrino magnetic moment search~\cite{note5}

\section{Conclusion}

Straightforward calculations indicate that one expects
thousands of coherent neutrino-nucleus interactions with recoil
energies $>10$ keV per ton of material per year of running at the SNS,
which is very promising.  Even few kilogram-scale experiments may have
measurable rates.  These estimates have been made for an experiment
with no background and no inefficiency; both will certainly be
important for a real experiment.  Sensitivities will need to be
reevaluated for a specific detector configuration for which
backgrounds and efficiencies can be estimated.  

Unambiguous detection
of the process is a first step; high statistics measurements will then
follow.  Such an experiment 
has significant potential for constraining NSI parameters;
magnetic moment and precision weak mixing angle measurements are also conceivable,
although pose a greater experimental challenge.

\begin{acknowledgments}

The author is grateful to D. Akerib, B. Balantekin, J. Beacom,
J. Collar, M. Dragowsky, Y. Efremenko, A. Friedland, H. Gao, C. Horowitz,
E. Kearns, G. McLaughlin, B. Mueller, R. Raghavan, T. Schutt,
C. Walter and A. Young, and especially 
J. Engel, C. Lunardini and D. McKinsey, for
comments and discussions.  C. Horowitz provided the form factors for
the calculation.  The author would also like to thank the Institute
for Nuclear Theory at the University of Washington, where this work
was started, for its hospitality.   The author's research activities
are supported by the Department of Energy and the National Science
Foundation.

\end{acknowledgments}

\bibliography{references}

\begin{thebibliography}{49}
\bibitem[{not({\natexlab{a}})}]{note1}
\bibinfo{note}{Note that for an SNS configuration, most sterile oscillation
  scenarios are already quite well constrained by other experiments, especially
  assuming a null result from Mini-BooNE.}
\bibitem[{not({\natexlab{b}})}]{note2}
\bibinfo{note}{One might conceivably eventually confront the current best
  $\varepsilon_{\mu \tau}$ limits with an SNS coherent elastic scattering
  experiment.}
\bibitem[{not({\natexlab{c}})}]{note3}
\bibinfo{note}{In principle one might do better with a spectral fit to separate
  $\nu_\mu$ and $\bar{\nu}_e$ contributions.}
\bibitem[{not({\natexlab{d}})}]{note4}
\bibinfo{note}{Note that one might achieve some cancellation of absolute
  source-flux-related systematic uncertainty by taking a ratio of prompt and
  delayed fluxes.}
\bibitem[{not({\natexlab{e}})}]{note5}
\bibinfo{note}{Note that a neutrino-electron magnetic scattering search can in
  principle be done at the SNS with low threshold detectors. The $\nu e$
  magnetic scattering cross-section is smaller by a factor of $Z^2$ than for
  neutrino-nucleus, but there are a factor of $Z$ more electron targets, so the
  signal is overall a factor of 10 smaller than for $\nu A$. This amounts to
  $\sim$10 $\nu e$ magnetic scattering events per ton per year at the SNS above
  10~keV, for $\mu_\nu$ at the current $\nu_\mu$ limit. The $\nu e$ magnetic
  signal has negligible SM background; however absolute rates are low for
  realistic target masses and likely to be subject to experimental
  backgrounds.}
\end{thebibliography}

\end{document}